\def\beq{\begin{eqnarray}}
\def\eeq{\end{eqnarray}}

\documentstyle[12pt,psfig,epsfig,graphicx,psfrag]{article}

\begin{document}
\begin{flushright}
LPT-ORSAY 01-116\\
NYU-TH/01/12/01\\
hep-ph/0112118\\
\end{flushright}
\vskip 1cm

\begin{center}
{\Large {\bf Dimming of  Supernovae by Photon-Pseudoscalar Conversion and 
the Intergalactic Plasma}}\\[1cm]
C\'edric Deffayet$^{a,}$\footnote{cjd2@physics.nyu.edu}, 
Diego Harari$^{b,}$\footnote{harari@df.uba.ar},
Jean--Philippe Uzan$^{c,d,}$\footnote{uzan@th.u-psud.fr} and
Matias Zaldarriaga$^{a,}$\footnote{mz31@nyu.edu}
\\
$^a${\it NYU Dept of Physics, 4 Washington Place, 
           New York, NY10003 (USA).}\\
$^b${\it Departamento de F{\'\i}sica, FCEyN - 
               Universidad de Buenos Aires,\\
          Ciudad Universitaria - Pab. 1, 1428 Buenos Aires (Argentina).}\\
$^{c}${\it Laboratoire de Physique Th\'eorique, UMR 8627 du CNRS, 
           Universit\'e Paris XI,\\ B\^at. 210 F--91405  
           Orsay Cedex (France).}\\
$^{d}${\it Institut d'Astrophysique de Paris, 98bis Bd. Arago,
           75014 Paris (France).}\\
\end{center}
\vskip 0.2cm

\noindent
\begin{center}
{\bf Abstract}
\end{center}

Recently Cs\'aki, Kaloper and Terning (hep-ph/0111311) suggested that
the observed dimming of distant type Ia supernovae may be a consequence
of mixing of the photons with very light axions. We
point out
that the effect of the plasma, in which the photons are
propagating, must 
be taken into account.
This effect changes the oscillation probability and renders the dimming
frequency-dependent, contrary to observations.  One may hope to
accommodate the data by averaging
the oscillations over many different coherence domains. We estimate the
effect of coherence loss, either due to the inhomogeneities of the
magnetic field or of the intergalactic plasma. These estimates indicate
that the achromaticity problem can be resolved only
with very specific, and probably unrealistic,  properties of the 
intergalactic medium.

\vskip 1cm

\section{Introduction}

The mixing between photons and axions, in an external magnetic field, is
a well studied mechanism~\cite{raffelt88,sikivie83,sikivie84,morris}
as is its analog with
gravitons~\cite{gertsenshtein62,zeldovich83}.  It is experimentally
used, since the pioneer work by Sikivie~\cite{sikivie83}, to 
constrain the
axion parameters~\cite{bibber89,hagmann98,kim98,sikivie97,gg} (see
also e.g. Ref. \cite{raffelt98} for an up to date review on such
experiments).  The astrophysical and cosmological implications of this
mechanism have also been studied~\cite{raffelt98}, and it was recently
advocated to be a possible explanation for the observed dimming of
distant type Ia supernovae~\cite{sn2,sn1} by Cs\'aki {\em et
al.}~\cite{Csaki:2001yk} (see also \cite{Erlich:2001iq} for the
analysis of supernovae data with the model
of~\cite{Csaki:2001yk}). The underlying idea is simply that the
luminosity of distant supernovae can be diminished due to the decay of
photons into very light pseudoscalar particles 
induced by intergalactic magnetic fields over
cosmological distances, and thus that the luminosity distance-redshift
relationship can mimic the one of a universe with a non zero
cosmological constant without need for a cosmological
constant. The pseudoscalar particles must have an electromagnetic
coupling similar to axions, and a specific and very small 
mass ($m\sim 10^{-16}$ eV) to avoid affecting the cosmic microwave
background anisotropy beyond its observed value.
Another aspect of photon-pseudoscalar conversion in
intergalactic
magnetic fields is the change of the polarization properties of
distant sources~\cite{diego} such as supernovae.

The implications on the cosmic microwave background of the similar
effect involving photon-graviton oscillation has been considered in
empty space~\cite{magueijo94,chen95} and it was shown that it becomes
negligible for standard cosmological magnetic fields~\cite{cillis96}
once the contribution of the intergalactic plasma is properly taken into
account (the case of axions is also considered in~\cite{du1}). The
effect of the inhomogeneities of the electron density upon the coherence
of the oscillations was also considered by Carlson {\em et
al.}~\cite{carl}.

The effect of this plasma was not addressed
in~\cite{Csaki:2001yk}, and it is our purpose to discuss
it in this work.

The paper is organized as follows: we first remind standard results on
photon-pseudoscalar oscillations in order to introduce our conventions
(see~\cite{raffelt88}, or~\cite{du1} where contributions from
Kaluza-Klein modes were also included). We then discuss specifically
the effect of the plasma for the parameters
relevant for type Ia supernovae.

\section{Photon--pseudoscalar mixing}\label{par5}

We consider the generic action for the  pseudoscalar-photon system 
\begin{eqnarray}\label{S4axion}  
  S_4&=&\int {\rm d}^4x\left[-\frac{1}{2}\left\lbrace  
  \partial^\mu a\partial_\mu a+m^2_{a}  
  a^2\right\rbrace +\frac{a}{M_a}F_{\mu\nu}  
     \widetilde F^{\mu\nu}-\frac{1}{4}F_{\mu\nu}F^{\mu\nu}\right],  
\end{eqnarray}  
where $a$ is the pseudoscalar field and $m_{a}$ its mass. $\widetilde
F_{\mu\nu}\equiv 
\epsilon_{\mu\nu\rho\sigma}F^{\rho\sigma}/2$ is the dual of the 
electromagnetic tensor, $\epsilon_{\mu\nu\rho\sigma}$ being the
completely antisymmetric tensor such that $\epsilon_{0123}=+1$. The
pseudoscalar couples to the photon with the coupling $1/M_a$.

We now consider an electromagnetic plane wave in the presence of a 
magnetic field $\vec B_0$ which is assumed constant on a  
characteristic scale $\Lambda_c$ in the sense that its variations  
in space and time are negligible on scales comparable 
to the photon wavelength and period. From the three independent vectors
\begin{equation}\label{base3d}  
  \vec e \equiv\frac{\vec k}{k}, \quad  
  \vec e_\parallel\equiv\frac{\vec B_{0,n}}{B_{0,n}},\quad  
  \vec e_\perp,  
\end{equation}  
we define a direct orthonormal basis of the three dimensional space
$(\vec e_\parallel,\vec e_\perp,\vec e)$. $\vec B_{0,n}$ is the
component of $\vec B_0$ perpendicular to the direction of
propagation $\vec k$.

The electromagnetic wave derives from a potential vector that can be
chosen to be of the form $\vec A=i(A_\parallel(s),A_\perp(s),0)
\hbox{e}^{-i\omega t}$ 
where $s$ is the coordinate along the direction of
propagation\footnote{with this convention the electric field of the
wave is simply $\vec{E} \equiv -\partial_t\vec A =(\omega
A_\parallel(s),\omega A_\perp(s),0)\hbox{e}^{-i\omega t} $. }.  With this
decomposition, the coupled Klein--Gordon and Maxwell equations derived
from the action (\ref{S4axion}) read
\begin{eqnarray}  
&&\left(\Box-m^2_{a}\right)a=\frac{4B_{0,n}}{M_a}  \nonumber
A_\parallel\\  
&&\Box A_\lambda=\frac{4B_{0,n}}{M_a}\omega\delta_{\lambda\parallel}a, 
\label{motion}
\end{eqnarray}
where $\lambda=\parallel,\perp$ denotes the polarization.
  
Assuming that the magnetic field varies in space on scales much larger
than the photon wavelength, we can perform the expansion
$\omega^2+\partial_s^2=(\omega+i\partial_s)(\omega-i\partial_s)
=(\omega+k)(\omega-i\partial_s)$ for a field propagating in the $+s$
direction. If we assume a general dispersion relation of the form
$\omega=nk$ and that the refractive index $n$ satisfies $|n-1|\ll1$,
we may approximate $\omega+k=2\omega$ and $k/\omega=1$.  This
approximation can be understood as a WKB limit where we set
$A(s)=|A(s)|\hbox{e}^{iks}$ and assumes that the amplitude $|A|$ varies
slowly, i.e. that $\partial_s |A|\ll k|A|$. In that limit, the above
system (\ref{motion}) reduces to the linearized system
\begin{equation}  \label{linsys}
\left(\omega-i\partial_s+{\cal M}\right)  
\left[  
\begin{array}{ccc}  
A_\perp\\A_\parallel\\a 
\end{array}  
\right]=0.  
\end{equation}  
The mixing matrix ${\cal M}$ is  defined by  
\begin{equation}  \label{mixmat}
{\cal M}\equiv\left(  
\begin{array}{cccc}  
\Delta_\perp&0\\
0&{\cal M}_\parallel
\end{array}  
\right)\qquad{\rm with}\qquad   
{\cal M}_\parallel\equiv\left(  
  \begin{array}{cc}  
    \Delta_\parallel&\Delta_M\\  
    \Delta_M&\Delta_m  
  \end{array}  
  \right).
\end{equation}  
The coefficients $\Delta_M$ and $\Delta_m$ are given by 
\begin{equation}  \label{eq12}
\Delta_M=\frac{2B_{0,n}}{M_a},  
\quad  
\Delta_m=-\frac{m^2_{a}}{2\omega}.
\end{equation}
The terms $\Delta_\lambda$ can be decomposed as
$\Delta_\lambda=\Delta_{\rm QED}+\Delta_{\rm CM}+\Delta_{\rm plasma}$.
The first term contains the effect of vacuum polarization giving a
refractive index to the photon (see e.g. Ref.~\cite{adler71}) and can
be computed by adding the Euler--Heisenberg effective Lagrangian which
is the lowest order term of the non--linearity of the Maxwell
equations in vacuum (see e.g.~\cite{euler36,iz}) to the action
(\ref{S4axion})\footnote{The equation of motion derived from (\ref{S4axion}) is 
(\ref{linsys}) with $\Delta_\lambda=0$. We intentionally omit the 
Euler--Heisenberg contribution in the presentation for the sake of 
clarity. Its Lagrangian is explicitly given by ${\cal L}_{EH}= 
\frac{\alpha^2}{90m_e^4}\left[(F^{\mu\nu}F_{\mu\nu})^2+\frac{7}{4} 
(F^{\mu\nu}\tilde F_{\mu\nu})^2\right]$.}.  The second term describes
the Cotton--Mouton effect, i.e. the birefringence of gases and liquids
in presence of a magnetic field and the third term the effect of the
plasma (since, in general, the photon does not propagate in
vacuum). Their explicit expressions are given by
\begin{eqnarray}\label{deltabis}  
  &&\Delta_{\rm QED}^\parallel=\frac{7}{2}\omega\xi,\quad  
  \Delta_{\rm QED}^\perp=2\omega\xi,\nonumber\\  
  &&\Delta_{\rm plasma}=-\frac{\omega_{\rm plasma}^2}{2\omega},\nonumber\\  
  &&\Delta_{\rm CM}^\parallel-\Delta_{\rm CM}^\perp=2\pi CB_0^2 
\end{eqnarray}  
with $\xi\equiv(\alpha/45\pi)(B_{0,n}/B_c)^2$, $B_c\equiv 
m_e^2/e=4.41\times10^{13}\,\hbox{G}$, $m_e$ the electron mass, $e$ the 
electron charge and $\alpha$ the fine structure constant.  $C$ is the 
Cotton--Mouton constant \cite{cotton}; its effect is to give only the 
difference of the refractive indices and the exact value of $C$ is 
hard to determine \cite{cotton2}.  The plasma 
frequency $\omega_{\rm plasma}$ is defined by 
\begin{equation}  
\omega_{\rm plasma}^2\equiv 4\pi\alpha\frac{n_e}{m_e},  
\end{equation}  
$n_e$ being the electron density. Note that  $\Delta_m$ is always negative whereas $\Delta_\lambda$ is positive if the  
contribution of the vacuum dominates and negative when the plasma  
term dominates.

As seen from Eq.~(\ref{linsys}), only the component $\parallel$,
i.e. parallel to the magnetic field, couples to the pseudoscalars, a first
consequence of which is that the polarization plane of a light beam
traveling in a magnetic field will rotate. 

The solution to the
equation of motion (\ref{linsys}) is obtained by diagonalizing ${\cal
M}_\parallel$ through a rotation
\begin{equation}\label{rotation}  
  \left[\begin{array}{c}A_\parallel'\\a'\end{array}\right]=  
  \left(\begin{array}{cc}\cos\vartheta&\sin\vartheta\\  
  -\sin\vartheta&\cos\vartheta\end{array}\right)  
  \left[\begin{array}{c}A_\parallel\\a\end{array}\right]  
\end{equation}  
with the {\it mixing angle} $\vartheta$ given by 
\begin{equation}\label{theta}  
 \tan2\vartheta\equiv2\frac{\Delta_M}{\Delta_\parallel-\Delta_m}. 
\end{equation}
By solving Eq.~(\ref{linsys}) in this new basis, one can easily
compute the probability of oscillation of a photon after a distance of
flight $s$ starting from the initial state
$(A_\parallel(0)=1,a(0)=0)$. It is explicitly given by
\begin{eqnarray}  
  P(\gamma\rightarrow a)\equiv\mid\langle A_\parallel(0)\mid a(s)  
  \rangle\mid^2 
  &=&\sin^2\left(2 \vartheta \right)\sin^2\left( 
  \frac{\Delta_{\rm osc}}{2}s\right),\label{p57}\\  
  &=&\left(\Delta_M s\right)^2 {\sin^2(\Delta_{\rm osc} s /2) 
  \over (\Delta_{\rm osc} s /2)^2}  
\end{eqnarray}  
with  the (reduced) oscillation wavenumber $\Delta_{\rm osc}$
 given by 
\begin{equation}  
\Delta_{\rm osc}=\frac{\Delta_\parallel-\Delta_m}{\cos 2\vartheta}  
=\frac{2\Delta_M}{\sin2\vartheta}.
\end{equation}  
The oscillation length is thus given by $\ell_{\rm
osc}\equiv2\pi/\Delta_{\rm osc}$.  We see that a complete transition
between a photon and a pseudoscalar is only possible when the mixing is
maximal ({\it strong mixing regime}) i.e. when $\vartheta\simeq\pi/4$.

\section{Application to Supernovae}

The quantities required for our discussion are $\Delta_M$,
$\Delta_m$,
$\Delta_{\rm plasma}$ and $\Delta_{\rm QED}$ respectively given by
equations (\ref{eq12}) and (\ref{deltabis}). It is useful to rewrite
them as
\begin{eqnarray}  
\frac{\Delta_M}{1\,{\rm cm}^{-1}}&=&  
       2\times10^{-26}\left(\frac{B_0}{10^{-9}\,\rm G}\right)  
       \left(\frac{M_a}{10^{11}\,\rm GeV}\right)^{-1},  \nonumber
\\  
\frac{\Delta_m}{1\,{\rm cm}^{-1}}&=&  
       -2.5 \times 10^{-28} \left(\frac{m_a}{10^{-16} 
        {\rm eV}}\right)^2 \left(\frac{\omega}{1 {\rm eV}} 
        \right)^{-1}, \nonumber
\\  
\frac{\Delta_{\rm plasma}}{1\,{\rm cm}^{-1}}&=&  
        -3.6\times10^{-24}\left(\frac{\omega}{1\,\rm eV}  
        \right)^{-1}\left(\frac{n_e}{10^{-7} \nonumber
        \,{\rm cm}^{-3}}\right),\label{plasmanumber}\\  
\frac{\Delta_{\rm QED}}{1\,{\rm cm}^{-1}}&=&  
        1.33\times10^{-45}\left(\frac{\omega}{1\,\rm eV}\right)  
        \left(\frac{B_0}{10^{-9}\,\rm G}\right)^2,  
\label{deltanum}  
\end{eqnarray}  
where we have used the facts that $1\,{\rm eV}\simeq 5\times  
10^{4}\,{\rm cm}^{-1}$, $1\,{\rm G}\simeq 1.95\times10^{-2}\,{\rm  
eV}^2$ in the natural Lorentz-Heaviside units where  
$\alpha=e^2/4\pi=1/137$.\\

The parameters chosen in \cite{Csaki:2001yk} are $M_a \sim 4 \times
10^{11} {\rm GeV}$, $m_a \sim 10^{-16} {\rm eV}$ and $B_0 \sim 10^{-9}
{\rm G}$.  The intergalactic medium (IGM) today is fully ionized, as
indicated by the lack of Gunn-Petterson effect \cite{gunn}.  Thus the mean
electronic density can be estimated to be (see e.g. \cite{peebles93})
\begin{equation}  
n_e\simeq10^{-7}\,{\rm cm}^{-3}.
\end{equation}
One immediately sees that, with this choice of parameters, $\Delta_{\rm
QED}$ is always negligible, whereas one has $|\Delta_{\rm plasma}|
\gg |\Delta_m|$ so that the plasma effects are always dominant over
the pure mass term of the pseudoscalar.\\

In order to be more specific, let us compare the mixing angles
$\vartheta$, oscillation wavenumbers $\Delta_{\rm osc}$ and
oscillation probabilities $P(\gamma \rightarrow a)$ with and without
including the effect of the intergalactic plasma, and for similar
choice of  parameters (we will use a subscript $_0$ for the values
computed without the plasma effect).

When one sets $\Delta_{\lambda}$ to zero in Eq.~(\ref{mixmat}), as
implicitly done in~\cite{Csaki:2001yk}, the mixing angle reduces to
\beq
\tan2\vartheta_0\equiv-2\frac{\Delta_M}{\Delta_m} \sim
40 \left(\frac{\omega}{1 {\rm eV}}  \right),  
\eeq
so that for optical photons (with $\omega \sim 1.5 $ eV to $3 $ eV) $ \tan2\vartheta_0 \gg 1$ and $ \vartheta_0 \sim
\pi/4$. This corresponds to a regime in which $\omega \gg m^2
/\Delta_M$ and the oscillation probability does not depend on $\omega$
so that the oscillation is {\it achromatic}.  The oscillation
wavenumber $\Delta_{\rm osc,0}$ is given then by
\beq
\Delta_{\rm osc,0} \sim 2 \Delta_M, 
\eeq
which is also independent of $\omega$.  With the choice of parameters
used in \cite{Csaki:2001yk}, $\Delta_{\rm osc,0}\simeq10^{-26}
{\rm cm^{-1}}$, so that the oscillation length is larger than the
size $s$ of the domain of coherence of the magnetic field considered
which is of order of a Mpc ($\sim 3 \times 10^{24} {\rm cm}$). The
probability of oscillation over a domain of size $s$ is then well
approximated by
\beq
P_0(\gamma \rightarrow a) \sim (\Delta_M s)^2, 
\eeq
which is of order $10^{-4}$. The number of such domains in our
Hubble radius $H_0^{-1}$, and on a given line of sight is given by
$H_0^{-1}/s$.  If one considers that the universe is made by patching
together such domains with uncorrelated $\vec{B}_0$, the coherence is
lost from domain to domain and one can simply sum up the probability of
conversion over each domain to obtain the probability of conversion of a
photon on cosmological distances given by (see also section 4)
\beq \label{Ptot}
P_{0,tot}(\gamma \rightarrow a) \sim \Delta_M^2 s H_0^{-1}.
\eeq
This number is of order 1, and one can thus expect a significant reduction
of the luminous flux over cosmological distances. This is the bottom
line of the mechanism proposed by Cs\'aki {\em et
al.}~\cite{Csaki:2001yk}.

Let us now include plasma effects. Since $|\Delta_{\rm plasma}|
\gg |\Delta_m|$, the mixing angle $\vartheta$ is now much smaller than
$\vartheta_0$ and
\beq
\vartheta \sim \frac{\Delta_M}{\Delta_{\rm plasma}}
\eeq 
of order $10^{-3}-10^{-2}$. With such a low mixing angle ({\it weak
mixing regime}), the probability of oscillation $P(\gamma \rightarrow
a)$ over a domain of size $s$ can be approximated by:
\beq \label{proba}
P(\gamma \rightarrow a) 
      &\sim&  (\Delta_M s)^2 \frac{{\rm sin}^2
      \left(\Delta_{\rm plasma} s /2\right)}{\left(\Delta_{\rm plasma} 
      s /2\right)^2} \nonumber \\
     &\sim& P_0(\gamma \rightarrow a)  
     \frac{ {\rm sin}^2\left(\Delta_{\rm plasma} 
     s /2\right)} {\left(\Delta_{\rm plasma} s /2\right)^2}.
\eeq
Notice then that the
oscillation wavenumber $\Delta_{\rm osc}$ is given by
\beq
\Delta_{\rm osc} \sim \Delta_{\rm plasma},
\eeq
so that the oscillation length is smaller than previously and is of the
same order as the size of the domain of coherence considered.  

\begin{figure} 
\psfrag{z}{{\footnotesize Redshift}} \psfrag{w}{{\footnotesize $\omega$ (eV)}}
\psfrag{P}{{\footnotesize $P(\gamma \rightarrow a)/ P_0(\gamma \rightarrow a)$ }}
\epsfig{file=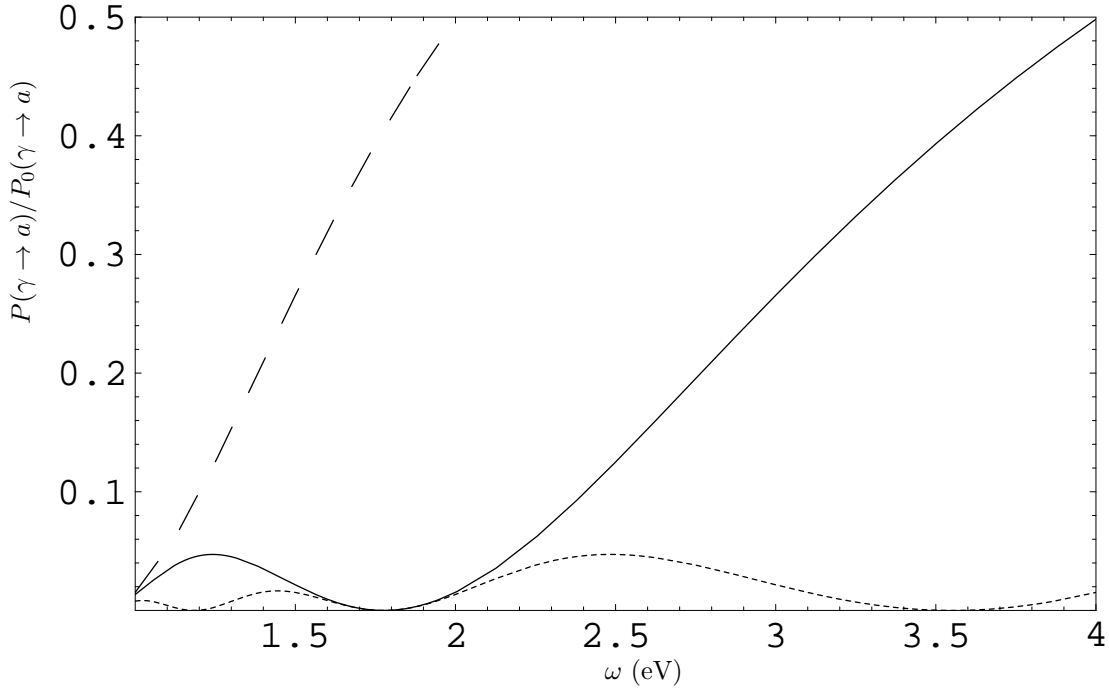, width=1.0\textwidth}
\caption{Ratio between the probability of oscillation of a photon 
into a pseudoscalar including the effect of the intergalactic plasma and
the
probability of oscillation when this effect is not considered (as done
in \cite{Csaki:2001yk}).  The curves are drawn as a function of the
photon energy,  for $M_a \sim 4 \times 10^{11} {\rm GeV}$, $m_a
\sim 10^{-16} {\rm eV}$ and $B_0 \sim 10^{-9} {\rm G}$, and for a
distance of flight $0.5$ Mpc (dashed line), $1$ Mpc (solid line) and $2$ Mpc (dotted
line). }\label{fig1}
\end{figure}

It follows that the probability of oscillation is lower than in the
previous case (with no plasma effects taken into account) and that it
is no longer achromatic (see Fig. \ref{fig1}). Supernovae observations not only argue for
a dimming of distant supernovae but also argue for an effect that is
achromatic, we will discuss this in more details in the next section. 

It is important to realize that when plasma effects are considered
there are two sources for the loss of coherence, spatial fluctuations
in the magnetic field and variations in the number density of
electrons (i.e. changes in the plasma frequency). For example, in the
case of photon-pseudoscalar conversion in the interstellar medium of our
galaxy \cite{carl}, the coherence length is most likely set by
the fluctuations in the plasma frequency. When this is the source of
coherence loss one expects generically that $s$ in equation
(\ref{proba}) also depends on frequency.

\section{Chromaticity constraints and coherence length}

As we mentioned in the previous section, once plasma effects are
considered the conversion probability can depend on photon
frequency. In this section we consider such constraints. 

Supernovae observations put a constraint on what is called 
the color excess between the $B$ and $V$ wavelength bands ($E[B-V]$). 
The color excess is defined as
\begin{equation} 
E[B-V]\equiv-2.5\ \ \log_{10}\left[{F^o(B) \over F^e(B)}\ \ {F^e(V) \over F^o(V)}\right],
\end{equation}
where $F^o$ (resp. $F^e$) is the observed (resp. emitted) flux and 
the $B$ (resp. $V$) band corresponds to $0.44 \mu {\rm m}$ (resp. $0.55 \mu
{\rm m}$). Observations constrain $E[B-V]$ to be lower than $0.03$  \cite{sn2}.
This can be translated to 
\begin{equation}
\label{chrom}
P(\gamma \rightarrow a)_V \left[ {P(\gamma \rightarrow a)_B\over P(\gamma
\rightarrow a)_V} -1 \right] < 0.03,   
\end{equation}  
or equivalently to the statement that the conversion probability has to scale with
photon wavelength weaker than $\lambda^{0.6}$ near the visual band.

We now consider different limits of equation (\ref{proba}) to
investigate the chromatic behavior. It is useful to introduce the
plasma length scale $\ell_p=\Delta_{\rm plasma}^{-1}$. We can rewrite (\ref{proba}) as
\beq \label{probb}
P(\gamma \rightarrow a) 
      &\sim&  (2 \Delta_M \ell_p)^2 {\rm sin}^2(s/2 \ell_p).
\eeq

In any astrophysically realistic situation,
the effective coherence length is either set
by the magnetic field or by the plasma frequency, so that it depends on 
properties of the IGM which are expected to have a significant 
dispersion\footnote{For example, in the model of Ref. \cite{loeb}, the
magnetic field in the intergalactic medium  was generated in quasars
and expelled in their outflows. The distribution of bubble sizes at
redshifts around zero is predicted to be extremely broad with two main 
components,  sizes
ranging from $0.5$ Mpc to about $5$ Mpc,  and a  mean size of order $1$
Mpc. The  
biggest bubbles have the largest magnetic field.}, 
and will then similarly exhibit such a dispersion.  \\

We start by considering the case in which the coherence length is set
by the magnetic field domains.  For simplicity we will take $s$ to be
distributed as a Gaussian variable with mean $s^*$ and dispersion
$\beta s^*$. The average of equation (\ref{probb}) over $s$ yields,
\begin{equation} 
\label{ave}
\left<P(\gamma \rightarrow a)\right> \sim 2(\Delta_M \ell_p)^2 
\left[ 1- \cos(s^*/\ell_p) e^{-(\beta s^*/\ell_p)^2/2}\right]
\end{equation}

In the limit $s^* \gg \ell_p$, it reduces to
\begin{equation}
\left<P(\gamma \rightarrow a)\right> \sim  2 (\Delta_M \ell_p)^2. 
\end{equation}
The plasma length $\ell_p$ scales with the photon frequency
so that the conversion probability is proportional to $\lambda^{-2}$
which is ruled out by the constraint in equation (\ref{chrom}).
Notice also that a conversion probability that
grows with frequency would produce a very significant
reddenning of sources located at distances such that the probability
reaches its saturation value for the shorter wavelengths. The absence
of such reddenning may provide even more stringent constraints on the
required achromaticity than that of Eq. (24). 

In the opposite limit, $s^* \ll  \ell_p$, we get 
\begin{equation} \label{Plim2}
\left<P(\gamma \rightarrow a)\right> \sim (\Delta_M s^*)^2, 
\end{equation}
which is achromatic if $s^*$ does not depend on frequency.  The total
probability of conversion will still be given by Eq.~(\ref{Ptot}) with $s^*$
replacing $s$, however since one must have $s^* \ll \ell_p$ for
(\ref{Plim2}) to hold, and since $\ell_p \sim 0.1 $ Mpc, this means that
the total probability of conversion will be lower than the one computed
in \cite{Csaki:2001yk} by at least a factor 10 to 100. One can try to
overcome this by changing the coupling of the pseudoscalar, it is however
difficult  because the coupling is already near the
astrophysical bound. Alternatively, the conversion probability may remain
achromatic as well as sufficiently large if the intergalactic magnetic
field were stronger in domains of size 
$s^*\ll\ell_p$. Faraday rotation measurements of distant quasars impose a
conservative bound $B_0~\sqrt{s_c} \le 10^{-9}~{\rm G}~\sqrt{\rm Mpc}$ on
the strength of an intergalactic magnetic field coherent over scales $s_c$
\cite{kronberg} (it may be somewhat 
stronger depending on its spatial structure \cite{farrar}).
The conversion probability could be achromatic in the visible band
and become of order unity at cosmological distances if the intergalactic
magnetic field had some very definite spectral properties,
for instance if $B_0$ were of order $10^{-8}$G over domains of average
size of the order of 10 kpc, and sufficiently weaker on larger domains. 
Additional constraints on the spatial distribution of the intergalactic
magnetic field compatible with the proposed mechanism arise from 
preventing excessive dispersion in the observed peak luminosity of
distant supernovae.
Clusters, for instance, may have magnetic fields significantly strong
and extended to make the conversion probability of order unity as photons 
get across them. 

Another possibility is  to consider the case
$s^* \sim \ell_p$ and to require that the average in equation (\ref{ave})
be achromatic enough. This usually does not happen, although it can be
accommodated by correctly choosing $\beta$ and $s^*$,
 or by a tight correlation between the strength of the magnetic field and 
the sizes of the domains, and requires very special, and likely unrealistic,
 statistical properties of the IGM\\

Let us now turn to the case where the coherence length is determined by the
spatial variations of the electron density. In this case the required achromaticity
would even be more of a fine tuning as the coherence length $\bar{s}$
will also depend on frequency. Just as in the case of the interstellar
medium studied in \cite{carl}, the frequency dependence will be set by
the clustering properties of the electron density.

In the case at hand we expect the intergalactic medium to be very
clumpy and complicated. Current numerical simulations indicate that
the baryons today can be found in several phases. About 30 \% by mass
is in a warm phase ($T\sim 5000 K$) that fills most of the volume of
the universe, about another $30 \%$ is in a warm-hot phase that
resides in non-virialized objects such a filaments and the rest
resides in virialized objects such as clusters and in condensed forms
such as stars and cool galactic gas (see e.g. \cite{dave} and references
therein).

The most relevant phase for our study is the warm one because it fills
most of the volume. It could be clumpy on scales smaller than
$\ell_p$, so the clumpiness of the IGM may be the most likely source
of coherence loss. However a detailed study of the loss of coherence
should probably involve studying lines of sight across these type of
simulations. It is important to realize that even in the limit where
the clumping scale, $L_c$, of the warm phase of the IGM  is smaller
than $\ell_p$ we still expect that $\bar{s}$ will depend on frequency. For
coherence to be lost, the random component of the accumulated phase of
the oscillation, $\phi$ , has to be of order one. The phase on each
segment of length $L_c$ is $L_c/\ell_p$ and accumulates over different
segments as a random walk, $\phi \sim L_c/\ell_p\ \sqrt{\bar{s}/L_c}\sim 1$.
In this limit, we estimate $\bar{s} \sim \ell_p^2/L_c$ which will again
induce a chromaticity that is ruled out by observations.\\

It seems that the only natural ways to avoid the chromaticity constraint
is (i) to assume that the magnetic field is responsible for setting the
coherence length of the oscillation and assume that $s^* \ll
\ell_p$, in which case either the coupling of the pseudoscalar needed to
accommodate the
dimming of the supernovae becomes uncomfortably large
or the magnetic field must have very definite strength and spectral
features, or (ii) to have very constrained properties of the IGM.

Finally, we note that the situation is not improved by giving a higher
mass to the pseudoscalar in order to have $\Delta_m > \Delta_{\rm
plasma}$; this
choice leads as well to chromaticity 
(since $\Delta_m$ and $ \Delta_{\rm plasma}$ have the same spectral dependence) 
and lowers the probability of oscillation (and the mixing angle). A last
logical possibility 
is that the mass of the pseudoscalar is such that $\Delta_m$ and $
\Delta_{\rm
plasma}$ are of the same order. In this case,  a strong mixing 
regime is possible whenever the electron density is such  that  
 $\Delta_m$ and $ \Delta_{\rm plasma}$ coincide with each other with 
accuracy $\Delta_M$. This can happen in the IGM from 
the statistical fluctuations in the electronic density. When this is the
case, a strong photon-pseudoscalar conversion can in principle take place
(this is
very analogous to the resonant MSW effect of neutrino physics). 
However, for the transition to be significant one has to maintain 
the resonant condition 
($\Delta_m \sim   \Delta_{\rm plasma}$ with accuracy $\Delta_M$) 
over a distance of the order of the oscillation length, 
$ \pi/ \Delta_M$, which given the numerical values (\ref{deltanum}) 
is also  very unlikely  (not to mention the fact that, a dimming 
of SNIa induced by such a resonant conversion would lead to a 
large dispersion in the observed SNIa magnitude).

\section{Conclusions}

We have shown that one can not ignore the effect of the intergalactic
plasma to derive how the luminosity of distant sources, such as
supernovae, is affected by a mixing with a hypothetical pseudoscalar
particle. In
most of the parameter space, this effect either renders the oscillation
frequency-dependent or lowers too much the oscillation probability.
There is a slight hope to accommodate the mechanism of
\cite{Csaki:2001yk}  if the IGM has very specific statistical properties.

\section*{Acknowledgments}
We thank  Gia Dvali, Andrei Gruzinov and  David Hogg,  for useful discussions.
The work of C.D. is sponsored in part by NSF Award PHY 9803174,
 and by  David and Lucile Packard Foundation
Fellowship 99-1462. The work of D.H. is supported by ANPCYT grant
03/05229.

\end{document}